\pdfoutput=1
\documentclass[preprints,article,accept,moreauthors,pdftex,10pt,a4paper]{mdpi}
\firstpage{1}
\pubvolume{xx}
\issuenum{1}
\articlenumber{1}
\pubyear{2018}
\copyrightyear{2018}
\history{Received: date; Accepted: date; Published: date}
\Title{The inverse-square interaction phase diagram: unitarity in the bosonic ground state}

\usepackage{amssymb}
\usepackage[english]{babel}
\usepackage{graphicx}
\usepackage{dcolumn}
\usepackage{bm}
\usepackage{color}
\Author{
G. E. Astrakharchik $^1$,
P. S. Kryuchkov $^2$,
I. L. Kurbakov $^2$,
Yu. E. Lozovik $^{2,3}$}
\address{
$^{1}$ \quad Departament de F\'{\i}sica, Universitat Polit\`{e}cnica de Catalunya, 08034 Barcelona, Spain\\
$^{2}$ \quad Institute of Spectroscopy (Russian Academy of Sciences), 108840, Troitsk, Moscow, Russia\\
$^{3}$ \quad National Research University Higher School of Economics, 109028 Moscow, Russia
}
\abstract{Ground-state properties of bosons interacting via inverse square potential (three dimensional Calogero-Sutherland model) are analyzed. A number of quantities scale with the density and can be naturally expressed in units of the Fermi energy and Fermi momentum multiplied by a dimensionless constant (Bertsch parameter). Two analytical approaches are developed: the Bogoliubov theory for weak and the harmonic approximation (HA) for strong interactions. Diffusion Monte Carlo method is used to obtain the ground-state properties in a non-perturbative manner. We report the dependence of the Bertsch parameter on the interaction strength and construct a Pad\'e approximant which fits the numerical data and reproduces correctly the asymptotic limits of weak and strong interactions. We find good agreement with beyond-mean field theory for the energy and the condensate fraction. The pair distribution function and the static structure factor are reported for a number of characteristic interactions. We demonstrate that the system experiences a gas-solid phase transition as a function of the dimensionless interaction strength. A peculiarity of the system is that by changing the density it is not possible to induce the phase transition. We show that the low-lying excitation spectrum contains plasmons in both phases, in agreement with the Bogoliubov and HA theories. Finally, we argue that this model can be interpreted as a realization of the unitary limit of a Bose system with the advantage that the system stays in the genuine ground state contrarily to the metastable state realized in experiments with short-range Bose gases.
}


\begin{document}
\tableofcontents

\section{Introduction}

The amazing progress in the field of ultracold Bose and Fermi atoms has provided a versatile tool for a highly controlled investigation of the properties of quantum systems.
The enormous advantage\cite{PitaevskiiStringariBook2016} of ultradilute quantum gases as compared to solid state systems and quantum liquids is the system high purity and controllability of the parameters, permitting one to tune the interaction strength by using a Feshbach resonance.
In particular, this permits one to revisit a number of old problems in new setups.
The Bardeen, Cooper, and Schrieffer theory\cite{Cooper56,BCS57,BCS57b} (Nobel prize in Physics in 1972), developed in the 60's to describe superconductivity in metals was more recently used to describe the BCS --- BEC crossover\cite{Grimm2004,Grimm04b,Ketterle04,Thomas04,Salomon04,GreinerRegalJin2005,Grimm07,Zwierlein12,InnsbruckFirstSoundAtUnitarity2013}.
In the original model the Cooper pairs are formed due to attraction between electrons and there is limited control over the interactions and the purity of the system. This is not the case in ultracold atoms where the Cooper pairs are formed between two spin components of alkali atoms.
Another characteristic example is that of a polaron\cite{LandauPekar48}, a quasiparticle introduced by Lev Landau to describe properties of an electron moving in a crystal, what was recently experimentally and theoretically studied in Bose\cite{Catani2012polaron1D,PhysRevLett.117.055302polaron2016,Cornell2016bosePolaron,Astrakharchik02c,Levinsen2015,Shchadilova2016,Grusdt2017,GrusdtAstrakharchikDemler2017,Volosniev2017,MassignanPRL2018} and Fermi\cite{ZwierleinPolaron2009,GrimmPolaron2012,Koschorreck2012polaron,Roati2017polaron} gases.
On the other hand, ultracold atoms also provided a way to experimentally address some problems which were otherwise unaccessible.
In particular, the {\em unitary} gas has been realized in a number of experiments\cite{Kinast05,SagiJin12,Hadzibabic17} and some of its properties are like those found in neutron stars\cite{PhysRevA.97.013601}.

The main idea behind universality is that two-body interactions in a dilute Fermi or Bose gas can be described by a single parameter\cite{Landau87}, namely the $s$-wave scattering length $a$, as the mean interparticle distance is large compared to the range of the interaction potential and its details are not important.
Under such conditions, the properties of the gas are governed by a single dimensionless quantity, the gas parameter $na^3$.
When this parameter is small, $na^3\ll 1$, usual perturbative methods can be applied\cite{LifshitzPitaevskiiBook,PitaevskiiStringariBook2016}.
In the opposite non-perturbative regime, $na^3\gg 1$, the $s$-wave scattering is large compared to the mean interparticle distance and it drops out of the considerations.
As a result, the only physically relevant scale which is left at zero temperature is the density.
Such conditions, known as a unitary regime, lead to a number of peculiar properties.
For example, the energy per particle can be written as $E/N = \xi E_F$ where $E_F$ is the Fermi energy, fully defined by its density, and $\xi$ is a dimensionless constant, often referred to as a Bertsch parameter.
The formal similarity between Bose and Fermi systems with density $n$ is that the only physical length scale is defined by the interparticle distance $\propto n^{-1/3}$, the momentum scale by the Fermi momentum $k_F \propto n^{1/3}$ and the energy unit is the Fermi energy $E_F = \hbar^2k_F^2/(2m) \propto \hbar^2n^{2/3}/m$.
As a result the scaling of the properties of a Bose system with density is similar to the scaling of an ideal Fermi gas.
Such a system will have thermodynamics similar to that of an ideal Fermi gas, even if there are strong interactions in the system and it lacks a perturbative expansion parameter.
Thus, knowledge of a single parameter $\xi$ provides a rather exhaustive description of the system, while its value cannot be found perturbatively.
The use of Quantum Monte Carlo methods has turned out to be very fruitful for the calculation of $\xi$ in two-component Fermi gases as such methods provide an {\em ab initio} non-perturbative approach (see Ref.~\cite{PhysRevA.87.023615} and references within for a historical overview).

The requirement for the unitary regime is that $R \ll l \ll a$, where $R$ denotes the range of the potential, $l \propto n^{-1/3}$ is the mean interparticle distance and $a$ the $s$-wave scattering length.
For purely repulsive interactions the $s$-wave scattering length is smaller than the range of the potential for any finite repulsion, $a<R$, and $a=R$ for the infinitely high interaction potential (hard spheres of diameter $R$).
It means that the condition $R \ll a$ cannot be satisfied for a repulsive interaction.
Instead, for an attractive interaction, the $s$-wave scattering can be large compared to the range of the potential.
For a zero-range potential, the energy of a two-body bound state is related to the $s$-wave scattering length as $E = -\hbar^2/(ma^2)$ where $m$ is mass of a single atom.
In the limit of a vanishing bound state, $E\to 0$, the $s$-wave scattering length diverges.
This means that the unitary regime essentially corresponds to the resonant scattering and it is experimentally created by tuning the magnetic field to the Feshbach resonance position.
This method works well for fermions where the attraction between different components is compensated by the quantum pressure originating from the Pauli exclusion principle.
The situation is drastically different for bosons where attraction between atoms leads to a many-body collapse (a bright soliton) which means that a homogeneous gas state is not thermodynamically stable.
Although, recently, a unitary Bose gas was observed in a number of experiments\cite{Hadzibabic2013BoseUNI,CornellJin014unitaryBosegas,Salomon2016uniBose}, it is not the true ground state but can probably be interpreted as a metastable state with a finite lifetime.
Moreover, the simple description in terms of a single parameter, the two-body $s$-wave scattering length, is not sufficient as Efimov three-body physics comes into play\cite{WernerKrauth2014} and it provides an additional length scale.
As a result, it has not been possible to create a ground-state unitary Bose gas.
From this perspective, it is interesting to investigate the possibility of using the inverse square pair interaction in order to realize the unitary regime in the ground state of a Bose system.
The key feature of the $1/r^2$ interaction is that it is scale free in a system composed of quantum particles.
The $s$-wave scattering length calculated for this interaction diverges, similar to what happens in the unitary regime for short-ranged gases.
As a result, there is no any other length scale except for the mean interparticle distance.
Consequently, the energy per particle can be expressed as $E/N = \xi E_F$ where the Bertsch parameter $\xi$ does not depend on the density, leading to the same thermodynamics as in an ideal Fermi gas with the Fermi energy $E_F$.

The effects of strong correlations eventually leading to the quantum crystallization typically can be understood from analysing the value of the dimensionless quantum parameter $q$, defined as the ratio of the characteristic interparticle interaction energy $V$ to the characteristic quantum kinetic energy $\hbar^2/ ml^2$, where $l$ is the average distance between particles. An important feature of the inverse-square interaction $Q\hbar^2 / mr^2$ is that the dimensionless quantum parameter $q \propto Q$ is independent of the interparticle distance or of the density $n$. We note that an analogous situation holds for electrons in graphene and in other Dirac materials. In these systems, due to the linear dependence of the kinetic energy from the momentum operator, the characteristic quantum kinetic energy $\propto \hbar V_F / r$ (where $V_F$ is the concentration-independent Fermi velocity, $V_F\approx c / 300$ with $c$ the speed of light). Therefore, for graphene and other Dirac materials, the characteristic quantum kinetic energy scales linearly with the average distance between electrons, similar to the linear scaling of the electron-electron Coulomb interaction $V(r) = e^2/r$. As a result, the governing quantum parameter is $q = e^2/(\hbar V_F)\approx 2$ for graphene. This leads, in particular, to the impossibility of Wigner crystallization for graphene (in the absence of strong magnetic fields).

There is a crucial difference between classical and quantum mechanical systems interacting with an attractive inverse-square potential, known as a {\em quantum anomaly}\cite{GuptaRajeev93,PhysRevLett.85.1590}.
While there are no qualitative changes in a classical system as $Q$ is changed, in a quantum system there exists a critical value of $Q=-1/4$.
For a weaker attraction, $Q>-1/4$, two interacting particles have a finite energy.
Instead, for more attractive interactions, $Q\le - 1/4$, the energy becomes infinite and the ground state ceases to exist, a phenomenon known as ``the fall of a particle to the center''\cite{Landau87}.
Suppression of the quantum collapse, i.e. of the non existence of the ground state has been studied for a central $1/r^2$  interaction potential in a number of systems by using mean-field methods\cite{PhysRevA.83.013607,PhysRevA.84.033616,PhysRevA.88.043638} and a quantum Monte Carlo technique\cite{AstrakharchikMalomed2015}.
Very recently the quantum anomaly has been observed in two-dimensional Fermi gases\cite{2018arXiv180308879H,2018quantumanomalyExp}.

The inverse-square potential is a famous example of exactly-solvable many body quantum systems in one dimension (see the book~\cite{Sutherlandbook} and references within).
Both the wave function and energy can be explicitly written as a function of the interaction parameter both in homogeneous and trapped geometries\cite{Calogero69b,Sutherland71,Sutherland71b,Sutherland72}.
It was observed that a similar construction of the ground state wave function in two dimensions leads to equivalent statistical averages as for a system of classical charges, while the quantum Hamiltonian in addition possesses three-body interaction terms\cite{Feigelman1997,Bardek2005,Feinberg2005,MozgunovFeigelman11}.

In the following we study the zero-temperature phase diagram of particles interacting via inverse square interactions and interpret the system properties in relation to the Bose system at unitarity.

\section{Homogeneous Calogero-Sutherland model}

We consider the following model Hamiltonian describing a system of $N$ particles of mass $m$ interacting via the repulsive inverse square interaction potential
\begin{eqnarray}
\hat{H}
=
-\frac{\hbar^2}{2m}\sum_{i=1}^{N}
\left(
\frac{\partial^2}{\partial x^2}
+\frac{\partial^2}{\partial y^2}
+\frac{\partial^2}{\partial z^2}
\right)
+\frac{\hbar^2}{m}\sum_{i>j}
\frac{Q}{(x_i-x_j)^2+(y_i-y_j)^2+(z_i-z_j)^2}\;.
\label{Eq:H}
\end{eqnarray}
In numerical simulations we impose periodic boundary conditions in a box of linear size $L$ so that the thermodynamic limit is obtained by increasing $N$ at a fixed density $n=N/L^3$. The dimensionless parameter $Q$ defines the strength of the interaction potential. A peculiarity of the Calogero-Sutherland interaction potential is that its dependence on the interaction distance, $r^{-2}$, scales similarly to the kinetic energy term. As a result, there is no other length scale than the density. This means that by changing the density it is not possible to change the phase of the system which instead is controlled by the dimensionless parameter $Q$. For small values of $Q$ it is possible to develop a perturbative theory of a weakly interacting Bose gas while in the opposite regime of a classical system with large $Q$ we use the harmonic crystal theory.

\section{Bogoliubov theory}

The scattering problem for the potential $Q = -1/8 + \nu^2/2$ results in a constant phase shift $\delta = -(\pi/2)(\nu+1/2)$, independent of the incident momentum $k$\cite{GuptaRajeev93}. In this sense, while the scattering phase is well defined, it is not possible to introduce the $s$-wave scattering length $a$, as the standard definition result in a divergent value, $a = - \lim_{k\to 0} \tan(\delta(k))/k \to \infty$. The independence of the phase shift $\delta$ of momentum $k$ is another manifestation of the impossibility to introduce a unit of scale from the two-body problem. In turn, this means that the standard perturbative theory developed for Bose gases is not applicable to the inverse square interaction and appropriate Bogoliubov theory should be developed.

In the limit of $Q\to 0$ the gas becomes weakly interacting. We develop Bogoliubov theory in this regime, based on the assumption that the condensate is macroscopically occupied. In the first quantization the Hamiltonian~(\ref{Eq:H}) is written as
\begin{eqnarray}
\hat{H} - \mu N
=
\int
\frac{\hbar^2}{2m}
\nabla\hat\Psi^\dagger({\bf r})
\nabla\hat\Psi({\bf r})
\;d{\bf r}
-\int \mu\hat\Psi^\dagger({\bf r})\hat\Psi({\bf r}) \;d{\bf r}
+
\frac{1}{2}
\int\!\!\!\!\int
\hat\Psi^\dagger({\bf r}) \hat\Psi^\dagger({\bf r'})
V({\bf r}, {\bf r'})
\hat\Psi({\bf r}) \hat\Psi({\bf r'})
\; d{\bf r} d{\bf r'}
\label{Eq:H:first quantization}
\end{eqnarray}
where $\hat\Psi({\bf r})$ and $\hat\Psi^\dagger({\bf r})$ are creation and annihilation the field operators. For particles obeying Bose-Einstein statistics, field operators satisfy the usual commutation relations
$[\hat\Psi({\bf r}), \hat\Psi^\dagger({\bf r})] = \delta({\bf r} - {\bf r'})$,
$[\hat\Psi({\bf r}), \hat\Psi({\bf r'})] = [\hat\Psi^\dagger({\bf r}), \hat\Psi^\dagger({\bf r'})] = 0$.

In order to calculate the Fourier transform of the interaction potential we first remove the short-range divergence of the potential by screening
\begin{eqnarray}
V^{scr}(r) = \frac{Q\hbar^2}{m(r^2+a^2)},
\label{Eq:V:screened}
\end{eqnarray}
where the screening length $a$ defines the value of the potential at zero according to $V^{scr}(r=0) = Q\hbar^2 / (ma^2)$. The Fourier transform of~(\ref{Eq:V:screened}) can be now evaluated
\begin{eqnarray}
V^{scr}_k = \int\limits_0^\infty V^{scr}(r) \frac{\sin (kr)}{kr} 4\pi r^2\;dr
 =\frac{2\pi^2\hbar^2Qe^{-ak}}{mk}
\label{Eq:Vk:screened}
\end{eqnarray}
and is related to the screened-Coulomb (Yukawa) function.

In a homogeneous system the field operators can be expanded in the basis of plane waves,
\begin{eqnarray}
\hat\Psi({\bf r}) = \sum_{\bf k} \hat a_{\bf k} \frac{1}{\sqrt{V}} e^{i {\bf k}\cdot{\bf r}},
\label{Eq:Psi}
\end{eqnarray}
where $\hat a_{\bf k}$ is an operator which annihilates a particle in the single-particle state with momentum ${\bf k}$. In a finite-sized box, only momenta satisfying the periodic boundary conditions are allowed. Substitution of~(\ref{Eq:Psi}) into~(\ref{Eq:H:first quantization}) gives the expression of the Hamiltonian in the second-quantization form
\begin{eqnarray}
\hat{H}-\mu N
=
\sum\limits_{\bf k}\left( \frac{\hbar^2k^2}{2m}\hat a^\dagger_{\bf k}a_{\bf k}  -\mu \right)
+
\frac{1}{2V}
\sum\limits_{{\bf k}_1, {\bf k}_2, {\bf q}}
V_{\bf q}
\hat a^\dagger_{{\bf k}_1+{\bf q}}
\hat a^\dagger_{{\bf k}_2-{\bf q}}
\hat a_{{\bf k}_1}
\hat a_{{\bf k}_2}.
\label{Eq:H:second quantization}
\end{eqnarray}

Within the Bogoliubov theory, it is assumed that as the ${\bf k} = 0$ state is macroscopically occupied, the corresponding operator can be treated as a number, $\hat a_{0} = \sqrt{N_0}$ and it can be used to construct a perturbative series,
\begin{eqnarray}
\hat{H}-\mu N
= \frac{N_0(N_0-1)}{2V}V_0
+
\sum\limits_{\bf k}\left( \frac{\hbar^2k^2}{2m}\hat a^\dagger_{\bf k}a_{\bf k}  -\mu \right)
+ \frac{N_0}{2V}\sum\limits_{\bf k}{}\!^{'}
\left[
V_{\bf k}(2\hat a^\dagger_{{\bf k}}\hat a_{{\bf k}}+ \hat a_{{\bf k}}\hat a_{{\bf -k}} + \hat a^\dagger_{{\bf k}}\hat a^\dagger_{{\bf -k}}
)
+2V_0\hat a^\dagger_{{\bf k}}\hat a_{{\bf k}}
\right].
\label{Eq:quadratic form}
\end{eqnarray}
Here and in the following the primed summations stand for summations over all $k$ except $k=0$.
In addition to the textbook expression\cite{LifshitzPitaevskiiBook,PitaevskiiStringariBook2016} derived for a short-range potential, here the Fourier transform of the interaction potential at finite momentum, $V_{\bf k}$, appears. Terms of a similar nature have appeared in Bogoliubov theory for dipolar interactions\cite{GhabourPelster14} which are also long-ranged.

A more important difference is that we do not have renormalization terms, which include correction beyond the lowest-order Born approximation to the coupling constant in terms of the $s$-wave scattering length to remove a divergence in the energy. The interaction potential in Eq.~(\ref{Eq:H}) would have a divergent $s$-wave scattering length and, indeed, is not a short-range potential. No second-order correction appears in our case and we immediately obtain convergent results for the Lee-Huang-Yang correction, as will be shown below.

Conceptually this is important, as there is a similarity between the $1/r^2$ interaction potential and the $\delta$-interacting potential in two dimensions as both scale as one over the distance squared.
The difference is that the theory with $\delta$-interaction potential suffers from the infrared divergency which is cured by the renormalization procedure which breaks the symmetry between classical and quantum systems.
This anomalous symmetry breaking known as the {\em conformal anomaly} was predicted for two-dimensional gases in Ref.~\cite{PitaevskiiRosch1997,OlshaniiPerrinLorent2010}
and was very recently experimentally observed in two experiments Ref.~\cite{2018arXiv180308879H,2018quantumanomalyExp}.
Thus, it is important to note that the $1/r^2$ interaction potential on the contrary does not have the conformal anomaly.

Quadratic in $\hat a_{\bf k}$ form~(\ref{Eq:quadratic form}) can be diagonalized using the Bogoliubov transformation,
\begin{eqnarray}
\hat a_{\bf k} &=& u_{\bf k} \hat b_{\bf k} + v_{\bf -k}^* \hat b^\dagger_{-\bf k},\\
\hat a^\dagger_{\bf k} &=& u_{\bf k}^* \hat b^\dagger_{\bf k} + v_{\bf -k} \hat b_{-\bf k}\;.
\label{Eq:Bogoliubov transformation}
\end{eqnarray}
By setting the coefficients in the off-diagonal terms to zero one obtains the Bogoliubov amplitudes
\begin{eqnarray}
u_{\bf k}^2 &=& \frac{1}{2}\left(\frac{\hbar^2k^2/(2m)-\mu+n_0(V_0+V_{\bf k})}{E({\bf k})}+1\right)\\
v_{\bf k}^2 &=& \frac{1}{2}\left(\frac{\hbar^2k^2/(2m)-\mu+n_0(V_0+V_{\bf k})}{E({\bf k})}-1\right)
\;,
\label{Eq:u,v}
\end{eqnarray}
where $n_0= N_0 / V$ is the density of the condensate. The Bogoliubov excitation spectrum is given by
\begin{eqnarray}
E({\bf k}) = \sqrt{\left({\frac{\hbar^2k^2}{2m}-\mu+ n_0V_0+n_0V_{\bf k}}\right)^2 - n_0^2 V^2_{\bf k}}
\;.
\label{Eq:Ebog}
\end{eqnarray}
Bogoliubov transformation~(\ref{Eq:Bogoliubov transformation}) is used to diagonalize Hamiltonian~(\ref{Eq:H:second quantization}) which leads to the diagonal form,
\begin{eqnarray}
\hat{H}-\mu N
= E_0 +
\sum\limits_{\bf k}{}\!^{'}
E({\bf k})\hat b^\dagger_{\bf k}\hat b_{\bf k}\;,
\label{Eq:diagonal}
\end{eqnarray}
where the ground-state energy
\begin{eqnarray}
E_0 = E_{MF} + E_{BMF}
\end{eqnarray}
has two contributions, the first coming from the mean-field interactions
\begin{eqnarray}
E_{MF}
= \frac{1}{2}N_0n_0V_0
\label{Eq:E:MF}
\end{eqnarray}
and the beyond-mean-field contribution stemming from quantum fluctuations (analogous to the Lee-Huang-Yang corrections for short-range interactions)
\begin{eqnarray}
E_{BMF} =
\frac{1}{2}\sum\limits_{\bf k}{}\!^{'}
\left[
E({\bf k})
-
\left(
\frac{\hbar^2k^2}{2m} - \mu + n_0(V_0+V_{\bf k})
\right)
\right]
\label{Eq:E:BMF}
\end{eqnarray}

For the inverse-square potential with no screening ($a=0$), the Fourier transform $V_{\bf k}$ for ${\bf k}= 0$ diverges in the thermodynamic limit, $N\to\infty$, resulting in an infinite energy per particle. At the same time, it is expected that the physical correlation functions must be well defined. The physically correct gapless spectrum is ensured by the following choice of the chemical potential
\begin{eqnarray}
\mu=n_0V_0\;,
\label{Eq:jellium}
\end{eqnarray}
which is consistent with the mean-field energy~(\ref{Eq:E:MF}). The resulting Bogoliubov spectrum,
\begin{eqnarray}
E(k) = \frac{\hbar^2n^{2/3}}{2m} \sqrt{8\pi^2Q \tilde k + \tilde k^4},
\label{Eq:Bogoliubov spectrum}
\end{eqnarray}
features a square-root dependence on the dimensionless momentum $\tilde k = k/n^{1/3}$ for low-energy excitations. Such excitations can be interpreted as {\em plasmons} similar to the Coulomb case and appear due to the long-range nature of the interactions. This situation should be contrasted with the linear phononic behavior typical for short-range potentials.

Quantum depletion of the condensate can be obtained by summation over the momentum distribution of particles excited out of the condensate. It follows from Eq.~(\ref{Eq:Bogoliubov transformation}) that the zero-temperature momentum distribution is proportional to the Bogoliubov amplitude~(\ref{Eq:u,v}) as $n_{\bf k} = \langle \hat a^\dagger_{\bf k} \hat a_{\bf k}\rangle = v_{\bf k}^2$. In the thermodynamic limit the sum can be approximated by an integral over the volume $v$,
\begin{eqnarray}
N_0 = N-
v\int\limits_0^\infty \frac{1}{2}\left(
\frac{\hbar^2k^2/(2m) + n V_{\bf k}}{E({\bf k})}-1
\right)
\frac{4\pi k^2\;dk}{(2\pi)^3}
= N \left(1 - \frac{Q}{3}\right)\;,
\label{Eq:No}
\end{eqnarray}
and the condensate fraction is suppressed linearly by the interaction strength $Q$ in the weakly interacting limit.

The quantum fluctuation correction~(\ref{Eq:E:BMF}) can be evaluated in the thermodynamic limit and is equal to
\begin{eqnarray}
E_{BMF} = -C_{BMF} N Q^{5/3} \frac{\hbar^2n^{2/3}}{m}\;,
\label{Eq:E:BMF:Q}
\end{eqnarray}
where $C_{BMF} = \frac{3}{5}\pi^{5/6} \Gamma(\frac{1}{3}) \Gamma(\frac{7}{6}) =
3.871$.
Remarkably, the $n^{2/3}$ density dependence of the bosonic energy~(\ref{Eq:E:BMF:Q}) is similar to that of an ideal Fermi gas.

\section{Quantum Monte Carlo approach}

Diffusion Monte Carlo method is used to study numerically the ground state properties. The statistical noise in the method can be greatly reduced by using a physically sound guiding wave function. Thus, before its construction we analyze in detail the long-and short-range behavior of the pair correlations in the system.

\subsection{Long-wavelength part of wave function}

Here we derive the long-wavelength limit of the many-body wave function following the recipe proposed by Reatto and Chester in Ref.~\cite{Reatto67} based on hydrodynamic grounds. Within the harmonic approximation, the long-range part of the many-body wave function can be written as
\begin{eqnarray}
\psi_{plasm}({\bf r}_1, \cdots, {\bf r}_N)
= \exp\left[
-\sum\limits_{k} \frac{m_k \omega_k}{2\hbar}\rho_k\rho_k^*
\label{Eq:plasmons}
\right]
\end{eqnarray}
where $m_k = m / (Nk^2)$ is the effective mass, $\omega_k$ is the long-wavelength plasmonic excitation spectrum and $\rho_{\bf k} = \sum\limits_{j=1}^{N} e^{i {\bf k \cdot r}_j}$ is the Fourier transform of the density operator. It means that the long-wavelength asymptote has a Bijl-Jastrow form,
\begin{eqnarray}
\psi_{plasm}({\bf r}_1, \cdots, {\bf r}_N)
= \exp\left[-\frac{1}{2} \sum\limits_{i<j} \chi(|{\bf r}_i - {\bf r}_j|)
\right],
\end{eqnarray}
where the correlation function is given by
\begin{eqnarray}
\chi(r) = \frac{1}{N}\sum\limits_{{\bf k}}\frac{m\omega_k}{2\hbar k^2} e^{i {\bf k}\cdot{\bf r}}.
\end{eqnarray}
The low-momentum excitation spectrum for the screened potential~(\ref{Eq:V:screened}) is obtained from the Bogoliubov excitation spectrum $\hbar\omega_k = \sqrt{n V_k \hbar^2k^2/m}$ with $V_k$ given by Eq.~(\ref{Eq:Vk:screened}), resulting in the thermodynamic limit result
\begin{eqnarray}
\chi(r) = \int\limits_0^\infty\frac{\sqrt{2Q}e^{-\frac{a k}{2}} \sin (k r)}{\pi r \sqrt{k n}}\;dk
=\frac{2\sqrt{Q}\sin\left[(1/2)\arctan\left(2r/a\right)\right]}{\sqrt{\pi}\sqrt{n} r \sqrt[4]{a^2+4 r^2}}\;.
\label{Eq:chi:screened}
\end{eqnarray}
In the limit of the unscreened potential, $a\to 0$, Eq.~(\ref{Eq:chi:screened}) takes the simple form
\begin{eqnarray}
\chi(r) = \sqrt{\frac{Q}{\pi n r^3}}.
\label{Eq:chi}
\end{eqnarray}
Interestingly, the asymptotic decay does not contain $\hbar$, as it does for the usual phononic systems, where
$\chi(r) = (mc) / (\pi^2\hbar n r^2)$.

\subsection{Short-range part of wave function}

When two particles approach each other, $r\to 0$, the divergence in the interaction potential implies that the wave function cannot diverge faster than $1/\sqrt{r}$ in order to have finite potential energy. For corresponding distances, contribution from other particles can be neglected resulting effectively in a two-body problem. The Schr\"odinger equation for two particles for the relative motion is
\begin{eqnarray}
-\frac{\hbar^2}{2m_r}\left[f''(r) + \frac{2}{r}f'(r)\right] + \lambda(\lambda+1)\frac{\hbar^2}{mr^2}f(r) = \frac{\hbar^2k^2}{2m_{r}} f(r)\;,
\label{Eq:2body}
\end{eqnarray}
where $r$ denotes the distance between the particles, $m_r = m/2$ is the reduced mass and we are searching for a spherical solution. For convenience we have expressed the interaction parameter as $Q = \lambda(\lambda+1)$. Equation~(\ref{Eq:2body}) can be explicitly solved for $k=0$ (zero-energy scattering problem) with the solution written as a linear combination of $r^\lambda$ and $r^{-1-\lambda}$ terms. The latter term leads to a diverging wave function and infinite potential energy, and is to be discarded. Thus, the scattering at zero momentum results in
\begin{eqnarray}
f(r) = C |r|^{\lambda},
\label{Eq:wf:short range}
\end{eqnarray}
with $C$ some normalization constant. Consequently, the Bijl-Jastrow terms should behave like ~(\ref{Eq:wf:short range}) for short distances.

In a many-body problem, the reduced mass $m_b$ used in Eq.~(\ref{Eq:2body}) to generate the variational wave function can be treated as a variational parameter\cite{Lutsyshyn2017} which parameterizes different variational states.
It was shown in Ref.~\cite{Lutsyshyn2017} that by doing so one obtains a better accuracy for the description of liquid helium which is a strongly correlated system. In our case the correlations are not so strong and we find that keeping the two-body value, $m_b=m/2$, is sufficient for performing calculations.

The scattering solution for zero momentum, $k=0$, is given by Eq.~(\ref{Eq:wf:short range}).
Taking into account the first correction due to finite value of $k>0$ we get the subleading correction,
\begin{eqnarray}
f(r) = C_1 |r|^{\lambda} + C_2 |r|^{\lambda+2}.
\label{Eq:wf:short range:c2}
\end{eqnarray}
The described procedure of searching for the solution in terms of a sum at short distances by cancelling the leading powers in the potential and kinetic energies is equivalent to the {\em cusp condition} commonly used in strongly diverging potentials (for example, Lennard-Jones one).

The cusp condition~(\ref{Eq:wf:short range:c2}) can be recovered in a formal way from the exact solution of the Schr\"odinger equation~(\ref{Eq:2body}) at a finite momentum,
\begin{eqnarray}
f(r) = C J_{\lambda}(kr),
\label{Eq:wf:short range:k}
\end{eqnarray}
in terms of the spherical Bessel function of the first kind $J_n(x)$.
Taylor expansion of Eq.~(\ref{Eq:wf:short range:k}) gives
\begin{eqnarray}
f(r) = C |r|^{\lambda}
\left(1
- \frac{(kr)^2}{4(\lambda+\frac{3}{2})}
+ \frac{(kr)^4}{32(\lambda+\frac{3}{2})(\lambda+\frac{5}{2})}
+ O((kr)^6)
\right),
\label{Eq:wf:short range:c4}
\end{eqnarray}
which, indeed, has the same functional form as the expansion~(\ref{Eq:wf:short range:c2}) obtained from the cusp condition.

\subsection{Guiding wave function}

In order to improve the convergence we use an importance sampling technique and we chose the guiding wave function in the Nosanow-Jastrow pair form,
\begin{equation}
\psi_T({\bf r}_1, \cdots, {\bf r}_N) =
\prod\limits_{i=1}^N e^{-\alpha({\bf r}_i - {\bf r}_j^c)^2}
\prod\limits_{i < j}f(|{\bf r}_i - {\bf r}_j|)\;,
\label{Eq:wf}
\end{equation}
where ${\bf r}_i^c$ denotes the radius vector of the each of the crystal lattice sites and $\alpha$ is a variational parameter governing particle localization near the crystal lattice site. When the localization is absent, $\alpha=0$, the wave function is translationally invariant and it describes the gas phase. Instead, for a finite localization strength, the wave function has broken translational symmetry and it is used to obtain properties in the solid phase.

We construct the Jastrow pair function $f(r)$ in such a way that it combines the physics of the two-body scattering at short distances and approaches Eq.~(\ref{Eq:wf:short range:c4}) as $r\to 0$, while at large distances it approaches the hydrodynamic asymptote given by Eq.~(\ref{Eq:chi}). We find it convenient to use the following choice of $f(r)$,
\begin{equation}
f(r) =
\left\{\begin{array}{cc}
c_1 r^\lambda(1 + c_2 r^2),&0<r\le R_{match}\\
c_3 \exp(-c_4/r^3 - c_5/ r^2),&R_{match}<r\le L/2\\
1,& r>L/2
\end{array}
\right.\;,
\label{Eq:Jastrow}
\end{equation}
where coefficients $c_1,\cdots,c_5$ are defined by the conditions of the continuity of $f(r)$ and its first derivative $f'(r)$ at the matching point $R_{match}$ and at $L/2$. The latter condition ensures that the guiding wave function~(\ref{Eq:wf}) satisfies the periodic boundary conditions imposed on the box of size $L\times L\times L$. The matching position $R_{match}$ is a free parameter and its value is optimized by minimizing the variational energy.

\section{Jellium model}

\subsection{Mean-field contribution}

While the beyond-mean-field contribution~(\ref{Eq:E:BMF}) to the energy is finite, the leading contribution~(\ref{Eq:E:MF}) is divergent. The reason is that the BMF term comes from well-behaved quantum fluctuations while the diverging term comes from the long-range contribution to the potential energy. This divergence is similar to that of a Coulomb gas, in which the divergence coming from $n_0V_0$ terms is removed by imposing the condition of the charge neutrality. A reasoning similar to that of a jellium model can be applied to our case. The energy of a uniform charge will be subtracted from the total and potential energies.

It is important to note that the use of the jellium model does not change the physical properties of the original Hamiltonian~(\ref{Eq:H}). All correlations in the system remain exactly the same. At the same time it becomes possible to quantify the potential energy which otherwise diverges in the thermodynamic limit.

The standard Bogoliubov theory does not distinguish the total $n$ and the condensate $n_0$ densities. That is, the subtraction of $n_0V_0$ from the chemical potential, appearing in the Bogoliubov theory~(\ref{Eq:jellium}), is actually equivalent to the subtraction of $nV_0$ in the jellium model.

\subsection{Direct summation}

In the jellium model for Coulomb charges, energy of the opposite uniform charge is subtracted from the energy of a finite system, Eq.~(\ref{Eq:jellium}). The simplest way to implement this condition is to truncate the interaction potential spherically to $V(r)=0$ for $r>L/2$. The finite size effects on the energy are significantly reduced by adding the missing {\it
tail} energy
\begin{eqnarray}
\frac{E_{tail}(n,L)}{N} = \frac{1}{2n} \int\limits_{L/2}^\infty V(r)g_2(r) 4\pi r^2 dr.
\label{Eq:Etail}
\end{eqnarray}
By assuming that the pair correlation function has reached its asymptotic value, $g_2(r) = n^2$, at half the size of the box
\begin{eqnarray}
\frac{E_{tail}(n,L)}{N}
= \frac{n}{2} \int\limits_{L/2}^\infty V(r) 4\pi r^2 dr\;.
\label{Eq:Etail2}
\end{eqnarray}
The implied condition of constant $g_2(r)$ is indeed satisfied in the gas phase as verified a posteriori in Fig.~\ref{Fig:g2}. Instead, in the crystal phase the self averaging over the peaks in $g_2(r)$ makes Eq.~(\ref{Eq:Etail2}) applicable for calculation of the mean energy correction. The jellium model assumes a uniform charge distribution, so that its contribution to the energy exactly coincides with the tail energy for distances $r>L/2$.

Instead, the contribution to the jellium energy coming from smaller distances definitely differs from the physical contribution at the same distances as it ignores correlations between particles. It can be explicitly evaluated resulting in
\begin{eqnarray}
\frac{E_{jel}}{N} = \frac{n}{2}\int\limits_0^{L/2} V(r) 4\pi r^2 dr = \pi Q N^{1/3}\frac{\hbar^2n^{2/3}}{m}\;.
\label{Eq:jellium:E}
\end{eqnarray}
Jellium contributions~(\ref{Eq:Etail2}-\ref{Eq:jellium:E}) will be subtracted from the total and potential energy eliminating the divergence in the thermodynamic limit. The jellium term~(\ref{Eq:jellium:E}) overestimates the potential energy which as it will be shown later is the dominant contribution to the total energy. As a result, the energy after the subtraction will be effectively negative while physically the total energy diverges to plus infinity in the thermodynamic limit.

\subsection{Smooth version of the long-range potential}

It is known that numerical simulations of systems with long-range interactions suffer from a number of technical issues. In particular, the following two challenges arise.

The first challenge to deal with is that the total energy per particle diverges in the thermodynamic limit. For example, the same problem arises in the calculation of the energy of an electron gas. Physically, the Coulomb energy of a same-charge system is diverging as the total charge becomes infinitely large in the thermodynamic limit. A standard procedure in this case consists in using the jellium model which adds the opposite charge uniformly distributed in the volume, so that the total charge remains equal to zero in the thermodynamic limit. In Monte Carlo simulations we subtract the jellium contribution~(\ref{Eq:jellium:E}) which permits us to calculate the energy in the thermodynamic limit. Within the Bogoliubov theory a gapless spectrum is recovered when a related term~(\ref{Eq:jellium}) is subtracted.

The second challenge is that once the diverging contribution is subtracted, the remaining integral converges slowly to the thermodynamic limit. A possible way out consists in using the Ewald summation technique\cite{Ewald1921,Osychenko2012ewald} which, essentially, converts a summation over images (slow power-law convergence) to a sum over coordinate and momentum space (Gaussian convergence). Instead, in the present work we use a method of ``smooth cutoff''\footnote{To be presented in details in a dedicated publication}. To do so, we substitute the original interaction potential $V(r)$ in Hamiltonian~(\ref{Eq:H}) by its ``smooth'' version
\begin{equation}\label{Vsm}
V_{\rm sm}(r)=V(r)\theta_{\rm sm}(r/R_c),
\end{equation}
where the smoothing function is taken as
\begin{equation}\label{thetasm}
\theta_{\rm sm}(x)=\left\{\begin{array}{cc}
(1-(2x-x^2)^{2k_1})^{k_2},&0\le x\le1\\0,&\mbox{otherwise}.
\end{array}\right.
\end{equation}
The ``tail'' energy corresponding to a uniform average for distances $r>R_c$ is calculated analytically resulting in the following contribution
\begin{equation}\label{Etail}
\frac{E_{\rm tail}}N=
\frac n2\int\limits_{-\infty}^{\infty}(V(r)-V_{\rm sm}(r))\;d{\bf r},
\end{equation}
which is added to the energy per particle.

There are three parameters which need to be specified. Two of them correspond to the shape of the smoothing function ($k_1$ and $k_2$) and the third one to the cutoff distance $R_c$. Parameters $k_1$ and $k_2$ should be large enough so that the cut-off method works effectively. In practice we find that the choices $k_1=6$ and $k_2=2$ provides good results. The cut-off parameter is taken as $R_c=0.73336L$ where $L$ is the linear size of the box side. The exact value of this parameter is calculated from the considering that the tail energy obtained with the smooth cutoff is equal to that of a sharp cutoff at $R_c = L / 2$, that is, $\lim_{A\to\infty} (\int_0^A (Q / r^2)(1-\theta_{\rm sm} (r / 0.73336L)) 4 \pi r^2dr-\int_ {L / 2}^A (Q / r ^ 2) 4 \pi r ^ 2dr) = 0$. The obtained number (0.73336) depends on the particular choice of the other parameters ($k_1 = 6 $ and $ k_2 = 2 $) and on the degree $\nu = 2$ of the power-law potential $V(r) = Q / r^{\nu} $. We have seen that the sharp cutoff for the gas gives a fast convergence of energy to the macroscopic limit.

To summarize, we use a smooth interaction potential~(\ref{thetasm}) which coincides with the original potential in the thermodynamic limit, but has a faster convergence to it.
In the following we will focus on the properties of the original Hamiltonian~(\ref{Eq:H}) in the $N\to\infty$ limit, obtained by the Monte Carlo method with the smooth interaction potential.

\section{Classical limit}

In the limit of large $Q$ the potential energy dominates and the properties of the system can be analyzed by using a semiclassical approach. The system crystalizes, as happens in classical systems at zero temperature.

\subsection{Equilibrium energy}

The main contribution to the energy comes from the potential energy which to leading order can be estimated by considering an ideal crystal energy,
\begin{eqnarray}
E_{crystal} = \frac{1}{2}\sum\limits_{{\bf R}, {\bf R'}} V({\bf R} - {\bf R'})
\label{Eq:E:classical}
\end{eqnarray}
where ${\bf R}$ denotes all equilibrium positions of the particles in the crystal (both within the same and other elementary cells). The energy per particle~(\ref{Eq:E:classical}) calculated with the $1/r^2$ interaction potential diverges leading to an infinitely large result in the thermodynamic limit. The jellium model can be conveniently used.

A possible way to do so is to limit summation in Eq.~(\ref{Eq:E:classical}) to a sphere of diameter $L$ and subtract the jellium contribution Eq.~(\ref{Eq:jellium:E}) from it. As a result, a finite classical energy is obtained.

The classical energy~(\ref{Eq:E:classical}) is slightly different for various possible crystal packings. The optimal one corresponds to face-centered cubic (fcc) packing with the energy
\begin{eqnarray}
\frac{E_{fcc}}{N} = -4.5372 Q \frac{\hbar^2n^2}{m},
\label{Eq:E:fcc}
\end{eqnarray}
which is only marginally below the energy $E_{bcc}/N = -4.5369Q\frac{\hbar^2n^2}{m}$ of the body-centered cubic lattice and $E_{hcp}/N = -4.5066Q\frac{\hbar^2n^2}{m}$ of the hexagonal close-packed lattice. Such tiny differences are hard to resolve in Monte Carlo calculations. In our simulations we always assume that the crystal packing is that of a fcc lattice as it is energetically preferable in the classical limit.

\subsection{Harmonic approximation}

The next correction to the energy and the excitation spectrum can be calculated by using a harmonic crystal theory\cite{ashcroft2011solid}. It is assumed that the particle positions ${\bf r}_i$ are close to the corresponding lattice sites ${\bf R}_i$ of the ideal crystal so that the ${\bf u}_i = {\bf R}_i - {\bf r}_i$ describes small deviations from it. The potential energy can be expanded to a quadratic form in which the constant contribution is given by Eq.~(\ref{Eq:E:classical}), the linear terms vanish as they should at in the minimum of the potential energy, and quadratic terms are described by the Hessian matrix $\Phi_{\mu\nu}({\bf R}) = (\partial^2 U({\bf R}) / \partial {\bf r}_\mu \partial {\bf r}_\nu)$, giving,
\begin{eqnarray}
E = E_{crystal}
+ \frac{1}{4}\sum\limits_{\substack{{\bf R}, {\bf R'}\\ \mu,\nu = x,y,z}}
[u_\mu({\bf R}) - u_\mu({\bf R'})]
\Phi_{\mu\nu}({\bf R} - {\bf R'})
[u_\nu({\bf R}) - u_\nu({\bf R'})]\;.
\label{Eq:E:HA}
\end{eqnarray}
The dynamic matrix
$D_{\mu\nu}({\bf R} - {\bf R'}) =
\delta({\bf R}-{\bf R'})
\sum\limits_{\bf R''} \Phi_{\mu\nu}({\bf R} - {\bf R''})
-\Phi_{\mu\nu}({\bf R} - {\bf R'})
$
can be conveniently introduced so that the energy can be interpreted as a quadratic form known as the harmonic approximation,
\begin{eqnarray}
E_{harm} =
\frac{1}{2}\sum\limits_{\substack{{\bf R}, {\bf R'}\\ \mu,\nu = x,y,z}}
u_\mu({\bf R})
D_{\mu\nu}({\bf R} - {\bf R'})
u_\nu({\bf R'})\;.
\label{Eq:E:HA2}
\end{eqnarray}

\subsection{Excitation spectrum}

The Newton equation of motion $m\ddot u = -\partial E_{harm} / \partial u$ can be solved according to the Bloch theorem as a wave with some dispersion law, $u({\bf r},t) = e^{i {\bf k}\cdot{\bf r} - \omega({\bf k}) t}$. The frequencies of the excitations can be found by diagonalizing the corresponding matrix. The correction to the leading term given by the equilibrium energy~(\ref{Eq:E:classical}) is then obtained by integrating the phonon energy $\hbar\omega({\bf k})$ over the first Brillouin zone. Here we limit ourselves to establishing the leading dependence on the interaction strength $Q$ in the energy and on momentum $k$ in the excitation spectrum.

Apart from some numerical coefficients, the Newton equation for the longitudinal mode can be recast in the following form
\begin{eqnarray}
m\omega^2(k)
= const
\sum\limits_{\bf R}(1-\cos({\bf k}\cdot{\bf R})) \left.\frac{\partial^2V(r)}{\partial r^2}\right|_{{\bf r}={\bf R}}.
\label{Eq:E:HA:sum}
\end{eqnarray}
The low-momentum limit $k\to 0$ corresponds to the summation over large distances. Even if the crystal has an irregular radial structure, for large distances it can be effectively neglected so that the sum can be approximated by an integral over the average density.
\begin{eqnarray}
m\omega^2(k)
\approx const
\int\limits_{R_{cutoff}}^{\infty} (1-\cos({\bf k}\cdot{\bf r})) \frac{\partial^2V(r)}{\partial r^2} d{\bf r},
\label{Eq:E:HA:int}
\end{eqnarray}
where $R_{cutoff}$, is some cutoff possibly needed to remove ultraviolet divergence (if present), that does not affect the low-momentum properties.

For the inverse square interaction, $V(r) = Q \hbar^2/(mr^2)$, this leads to $m\omega^2(k) = const Q\hbar^2|k|/m$ and the low-lying excitation spectrum is
\begin{eqnarray}
E(k) = \hbar\omega(k)
\propto \sqrt{Q} \frac{\hbar^2n^{2/3}}{m} \left(\frac{|k|}{n^{1/3}}\right)^{1/2}.
\label{Eq:E:HA:E(k)}
\end{eqnarray}
A number of important properties can be noted: (i) the low-lying excitations are not linear in the momentum but rather follow a square-root dependence (ii) the strength of the excitation scales as $\sqrt{Q}$ with the interaction parameter (iii) the energy can be expressed in terms of the Fermi energy and $k$ in Fermi momentum. Interestingly, properties (i---iii) found here in a classical system also hold true in a quantum weakly-interacting Bose gas, as given by the Bogoliubov spectrum~(\ref{Eq:Bogoliubov spectrum}). It is important to note that Planck's constant appearing as $\hbar^2$ in the excitation spectrum of a classical system has no profound meaning. Indeed, according to Eq.~(\ref{Eq:E:HA:int}) the frequency is proportional to the square root of the interaction potential which already contains $\hbar^2$. The conversion of the frequency $\omega$ to the energy $E(k)$ requires mutliplication by another $\hbar$. In this way the result contains the square of the Planck's constant although the underlying processes are entirely classical.

It is also interesting to check what is the dependence of the low-lying excitation spectrum~(\ref{Eq:E:HA:E(k)}) on momentum for other power-law potentials. It can be explicitly checked, that Coulomb $1/r$ charges in two dimensions also follow the law, $E\propto \sqrt{Q|k|}$. In this sense we refer to the low-lying excitations as {\em plasmons}. Instead, $1/r^3$ interaction in 3D is similar to one-dimensional chain of Coulomb $1/r$ charges featuring a ``weak'' logarithmic prefactor in front of the linear dispersion term\cite{Schulz93,AstrakharchikGirardeau2011,Ferre2015}, $E(k) \propto \sqrt{|\ln(k)||k|}$, being still a long-range potential. Instead, for the short-range potentials ($1/r^4$, $1/r^5$, etc) the dispersion relation obtained with Eq.~(\ref{Eq:E:HA:E(k)}) is linear.

Also, we note that from Eq.~(\ref{Eq:E:HA:E(k)}) it is possible to conclude that the leading correction to the ground-state energy scales as $\sqrt{Q}$ and is small compared to the leading term~(\ref{Eq:E:classical}) linear in $Q$.

\section{Numerical results}

\subsection{Thermodynamic properties}

\begin{figure}[ht]
\begin{center}
\vskip 0 pt \includegraphics[clip,width=0.6\columnwidth]{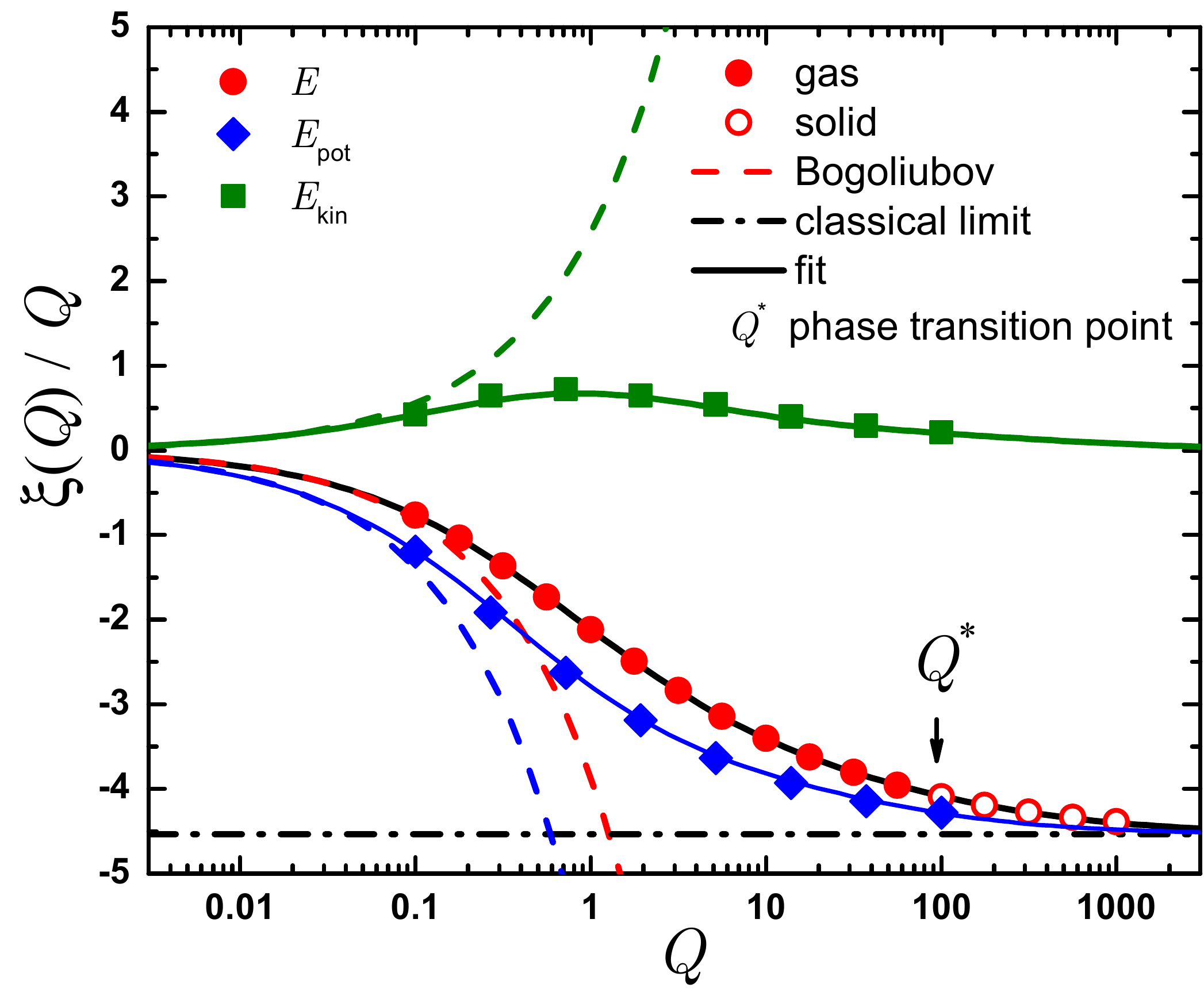}
\end{center}
\caption{
Bertsch parameter $\xi(Q)$ extracted from the ground-state energy $E$ scaled by $NQ$ as a function of $Q$ on a semi-logarithmic scale.
Symbols are Monte Carlo results,
solids lines are fits;
dashed lines are asymptotic expansions.
Including,
solid circles, total energy in the gas phase;
open circles, total energy in the solid phase;
diamonds, potential energy;
squares, kinetic energy;
dashed lines, Bogoliubov theory prediction valid for small $Q$
for the total~(\ref{Eq:E:BMF:Q}),
kinetic~(\ref{Eq:E:BMF:Ekin}),
and potential~(\ref{Eq:E:BMF:Epot}) energies;
dash-dotted line, large $Q$ limit of a classical fcc crystal~(\ref{Eq:E:fcc})
solid lines, fits given by Eqs.~(\ref{Eq:fit}) for the total energy,
Eq.~(\ref{Eq:scaling:Epot}) for the potential energy
and Eq.~(\ref{Eq:scaling:Ekin}) for the kinetic energy.
The arrow shows the position of the gas-solid phase transition $Q^*$.
The data points represent the energy of a gas for $Q<Q^*$ and of a solid otherwise.
Note that the jellium contribution is subtracted from the total and potential energies resulting in a negative energy even if the interaction is entirely repulsive.
}
\label{Fig:E}
\end{figure}

One of the main results of the present work is the prediction for the dependence of the Bertsch parameter $\xi$ on the interaction strength $Q$. In practice we extract $\xi(Q)$ from the energy per particle obtained by diffusion Monte Carlo calculations and extrapolated to the thermodynamic limit. For large $Q$ the Bertsch parameter grows linearly with $Q$ and it is graphically more convenient to present $\xi(Q)/Q$ ratio instead. This scaled quantity is reported in Fig.~\ref{Fig:E}. Note that the diverging mean-field (jellium) contribution~(\ref{Eq:jellium:E}) is subtracted. In this way we obtain an extensive quantity which is additive with the system volume. Due to this substraction, the resulting energy is negative even if it physically corresponds to a purely repulsive system.

The gas properties in the weakly interacting regime, $Q\to 0$, are correctly described by the Bogoliubov theory, as shown by a dashed line in Fig.~\ref{Fig:E}. We find that for $Q\approx 0.1$ the predictions of the Bogoliubov theory are very close to the DMC results and we conclude that the Bogoliubov theory can be safely used for obtaining the energy at smaller values of $Q$. The beyond mean-field correction~(\ref{Eq:E:BMF:Q}) scales as $\xi(Q)\propto Q^{5/3}$ and originates from the quantum fluctuations. It is interesting to note that the corresponding correction for a short-range Bose gas, given by the Lee-Huang-Yang term\cite{Huang57,Lee57,Lee57b}, is positive while we find a negative correction. The reason for this is that in the second order theory for short-range interaction one needs to take into account the renormalization of the coupling constant as opposed to the first Born approximation. Instead, the inverse-square potential corresponds to an infinite $s$-wave scattering length which does not need to be renormalized. The second-order perturbative theory lowers the energy as reflected by the negative sign of the correction.

As $Q$ is increases, the interactions become stronger and $\xi(Q)$ increases in absolute terms. As $Q$ reaches as certain value, $Q^*$, a phase transition from gas to a solid occurs. The exact position of this transition will be commented later. For ever stronger interactions the potential energy becomes much larger than the kinetic one and in the limit $Q\to\infty$ the energy of a classical crystal~(\ref{Eq:E:fcc}) is recovered.

While $\xi(Q)$ experiences a kink at the critical value, $Q=Q^*$, the discontinuity in the first derivative can be hardly perceived in the curve shown in Fig.~\ref{Fig:E}. We find that the following Pad\'e approximant
\begin{eqnarray}
\xi^{fit}(Q) = Q \frac{a_1Q^{2/3} + a_2Q^{4/3} + a_3Q^{2}}{b_0 + b_1Q^{2/3} + b_2Q^{4/3} + b_3Q^{2}}\;,
\label{Eq:fit}
\end{eqnarray}
accurately describes the thermodynamic dependence of $\xi$ where the coefficients $b_0 = 0.026; b_1 = 0.587; b_2 = 0.682; b_3 = 0.0345; a_2 = -2.557$ are obtained from the fit and conditions $a_1 = -C_{BMF} b_0$, $a_3 = E_C b_3$ are imposed to reproduce the BMF expansion~(\ref{Eq:E:BMF:Q}) and the energy of a classical crystal~(\ref{Eq:E:fcc}) in the corresponding limits.

\subsection{Quantum phase transition}

\begin{figure}[ht]
\begin{center}
\vskip 0 pt \includegraphics[clip,width=0.6\columnwidth]{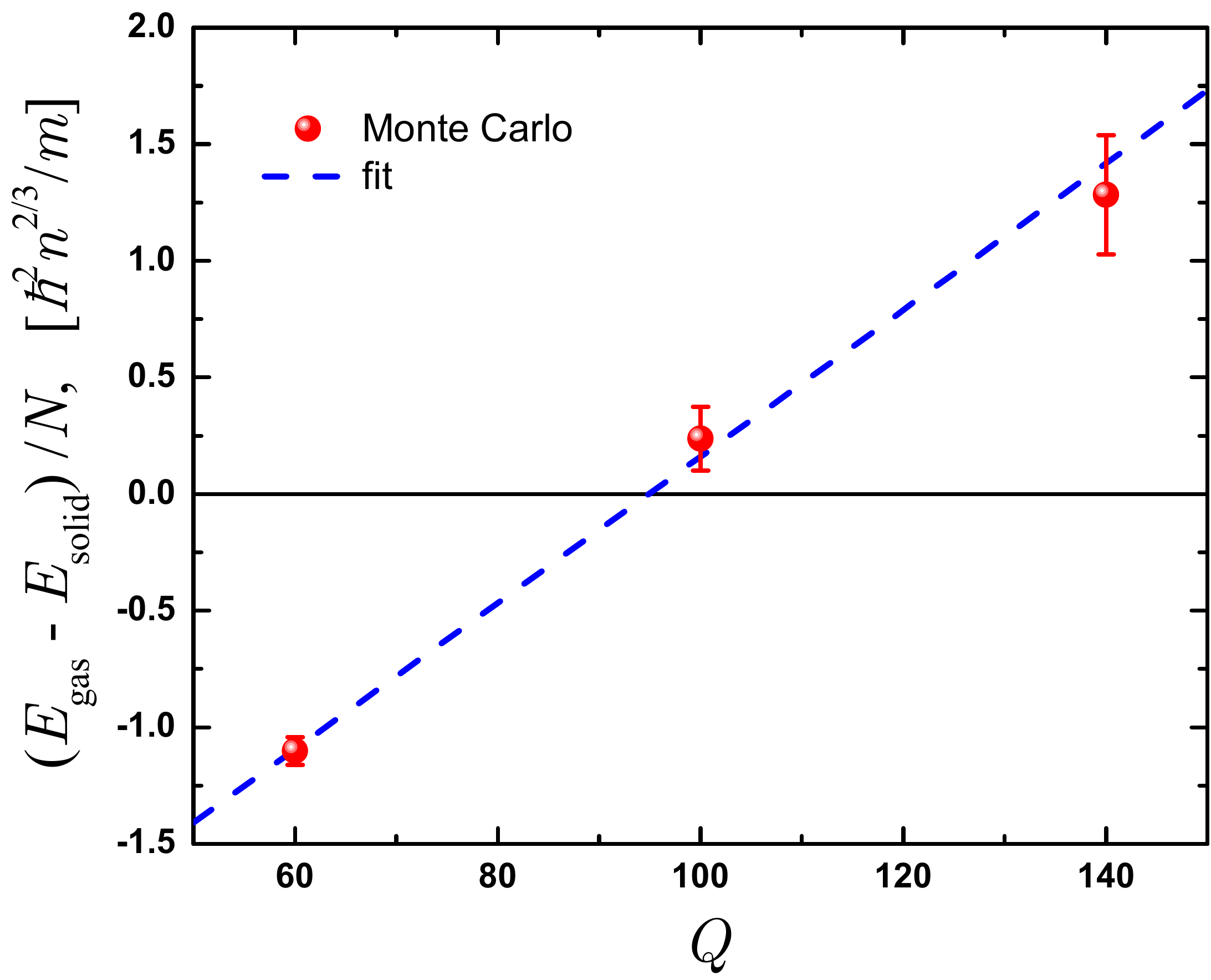}
\end{center}
\caption{
The difference of the energy per particle between the gas and solid phases as a function of $Q$.
Symbols report Monte Carlo data,
line shows the linear fit.
The point where the difference is equal to zero corresponds to the location of the phase transition.
}
\label{Fig:Phase}
\end{figure}

The second main prediction of our work is the position of the gas-solid phase transition. In order to find its location we calculate the energy in the gas $E_{gas}$ and in the solid $E_{solid}$ phases using guiding wave functions~(\ref{Eq:wf}) of appropriate translational symmetries. For each of the energies the thermodynamic value is obtained by using a quadratic fit in powers of $1/N$ for the energy with the jellium contribution subtracted. The difference between the energies in the gas and the solid phases is shown in Fig.~\ref{Fig:Phase}. For small values of $Q$ the gas phase is energetically preferable while for large $Q$ the crystal phase is the ground state. The point where the difference is equal to zero corresponds to the phase transition point. We estimate its location as $Q^* = 94(5)$.

It is important to note that no Maxwell double-tangent construction should be used in the present case as the phase transition is not caused by a change in the density or pressure but rather from change of an external parameter, $Q$. Indeed, by changing the density it not possible to provoke the phase transition and the phase always remains the same due to the scaling property. Thus, the zero-temperature phase transition is generated by changing parameter $Q$ in Hamiltonian~(\ref{Eq:H:first quantization}) rather than keeping the Hamiltonian and changing the volume or the number of particles. To a certain extent this is reminiscent of a situation where the phase transition is caused by a magnetic field which is an external parameter.

\subsection{Coherence and structural properties.}

\begin{figure}[ht]
\begin{center}
\vskip 0 pt \includegraphics[clip,width=0.6\columnwidth]{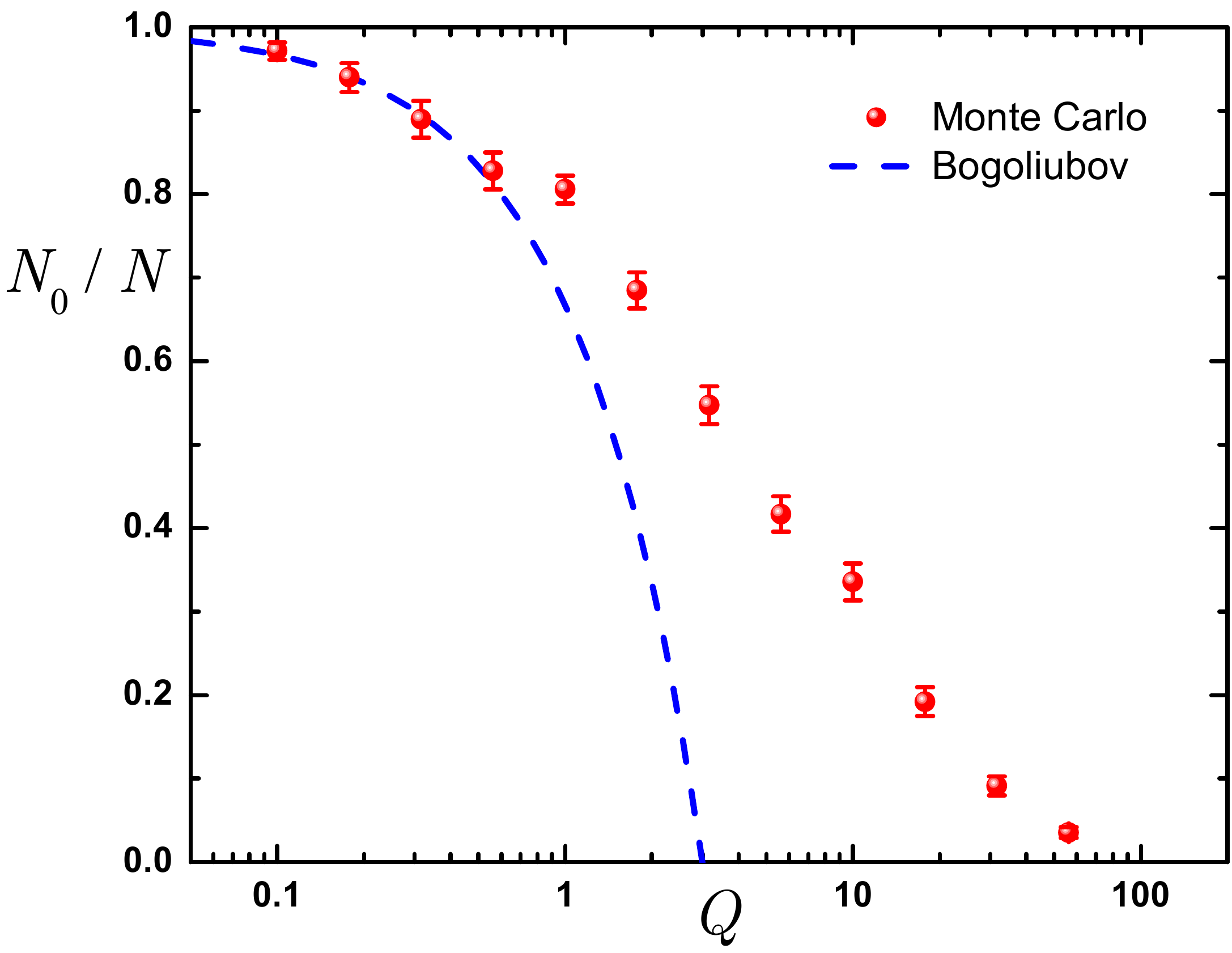}
\end{center}
\caption{
Condensate fraction as a function of $Q$.
Symbols report Monte Carlo data;
dashed line shows the Bogoliubov theory prediction given by Eq.~(\ref{Eq:No}).
}
\label{Fig:CF}
\end{figure}

Bose-Einstein condensation happens at low temperature in the gas phase. In order to quantify it and validate the use of the Bogoliubov theory we calculate the condensate fraction $N_0/N$. We calculate the large distance asymptote of the one-body density matrix and use an extrapolation procedure from the variational and diffusion Monte Carlo estimator to minimize a possible residual dependence on the guiding wave function. The dependence of $N_0/N$ on the interaction parameter $Q$ in the thermodynamic limit is reported in Fig.~\ref{Fig:CF}. The condensate fraction is close to unity for small values of $Q$ showing that the condensation is complete. The departure from this value is correctly captured by the perturbative Bogoliubov theory shown with a dashed line in Fig.~\ref{Fig:CF}. The condensate fraction decreases monotonically as $Q$ is increased and becomes very small close to the phase transition point.

\begin{figure}[ht]
\begin{center}
\vskip 0 pt \includegraphics[clip,width=0.6\columnwidth]{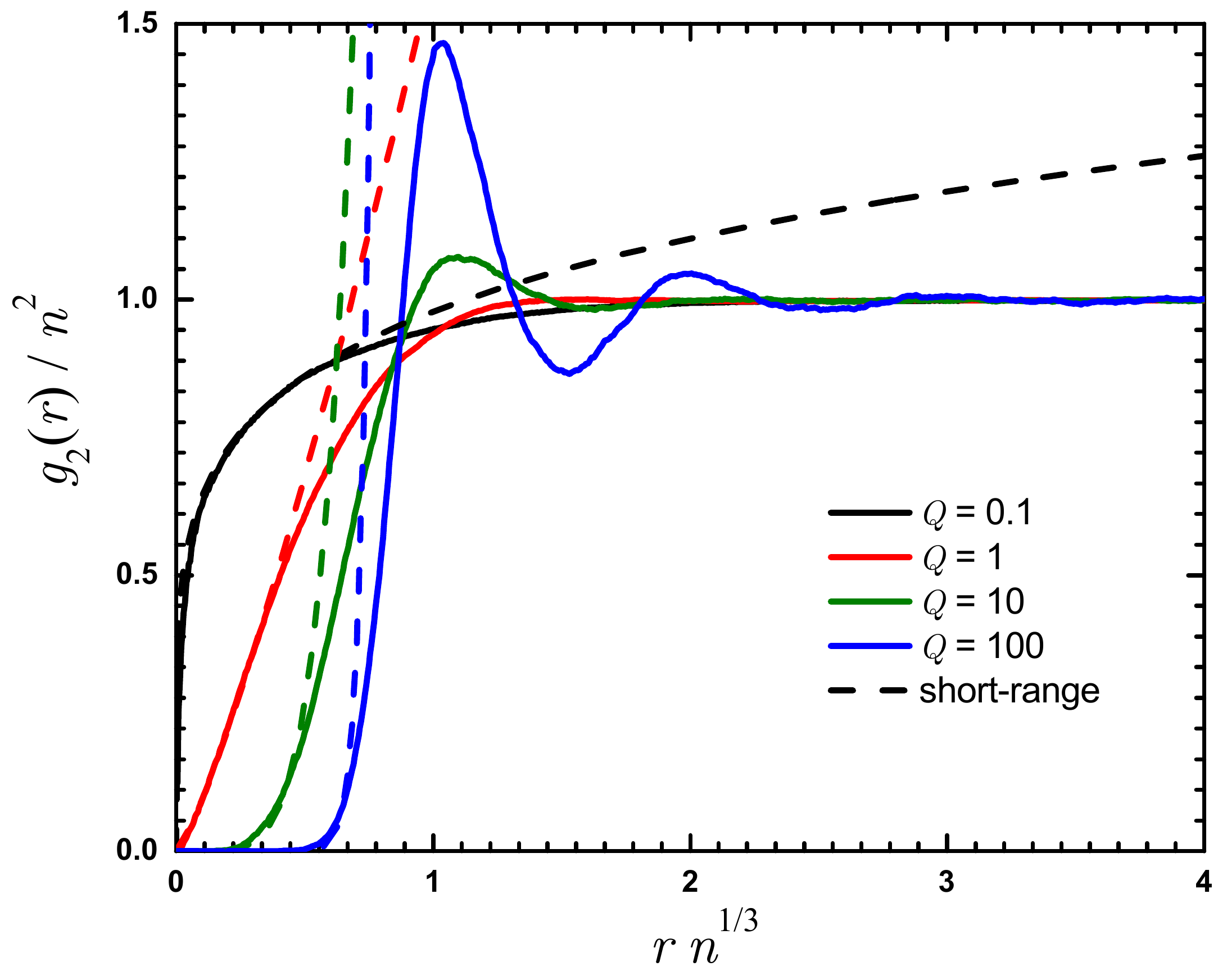}
\end{center}
\caption{
Pair distribution function in the gas phase for different values of $Q=0.1; 1; 10; 100$ (from top to bottom).
Solid lines, Monte Carlo results;
dashed lines, short-range expansion $|r|^{2\lambda}$ with the coefficient of proportionality fixed by a fitting procedure (see Eq.~(\ref{Eq:wf:short range:c4})).
Note that the point with $Q=100$ corresponds to a metastable state, as the ground state in this regime is a solid.
}
\label{Fig:g2}
\end{figure}

In order to study how the spatial correlations evolve with $Q$ we measure the pair distribution function $g_2({\bf r}, {\bf r'}) = \langle n({\bf r})n({\bf r'})\rangle$ which quantifies the density-density correlations in the system. In the gas phase it is isotropic and depends on the absolute value of the distance between two points, $|{\bf r}-{\bf r'}|$. A number of characteristic examples for the gas phase are reported in Fig.~\ref{Fig:g2}. When the distance between two particles is small, ${\bf r}-{\bf r'}\to 0$, the diverging interaction potential defines the wave function $f(r)$ according to Eq.~(\ref{Eq:wf:short range:c4}). For $\lambda>0$ the wave function vanishes at the contact resulting in $g_2(r=0)=0$. We verify that the $g_2(r) \propto f^2(r) \propto |r|^{2\lambda}$ for small distances and arbitrary interaction strength $Q$. Generally, this polytropic behavior is non-analytic unless the interaction parameter corresponds to the even powers of $\lambda$. A similar situation happens in a one-dimensional Calogero-Sutherland model\cite{Astrakharchik06c}.

The pair distribution function shows a smooth monotonic behavior in the Bogoliubov limit of small $Q$, typical for weakly interacting Bose gases. As $Q$ is increased, correlations becomes stronger and peaks start being formed. Close to the transition point, the height of the peak is around $1.5$ in units of the average density.

\begin{figure}[ht]
\begin{center}
\vskip 0 pt \includegraphics[clip,width=0.6\columnwidth]{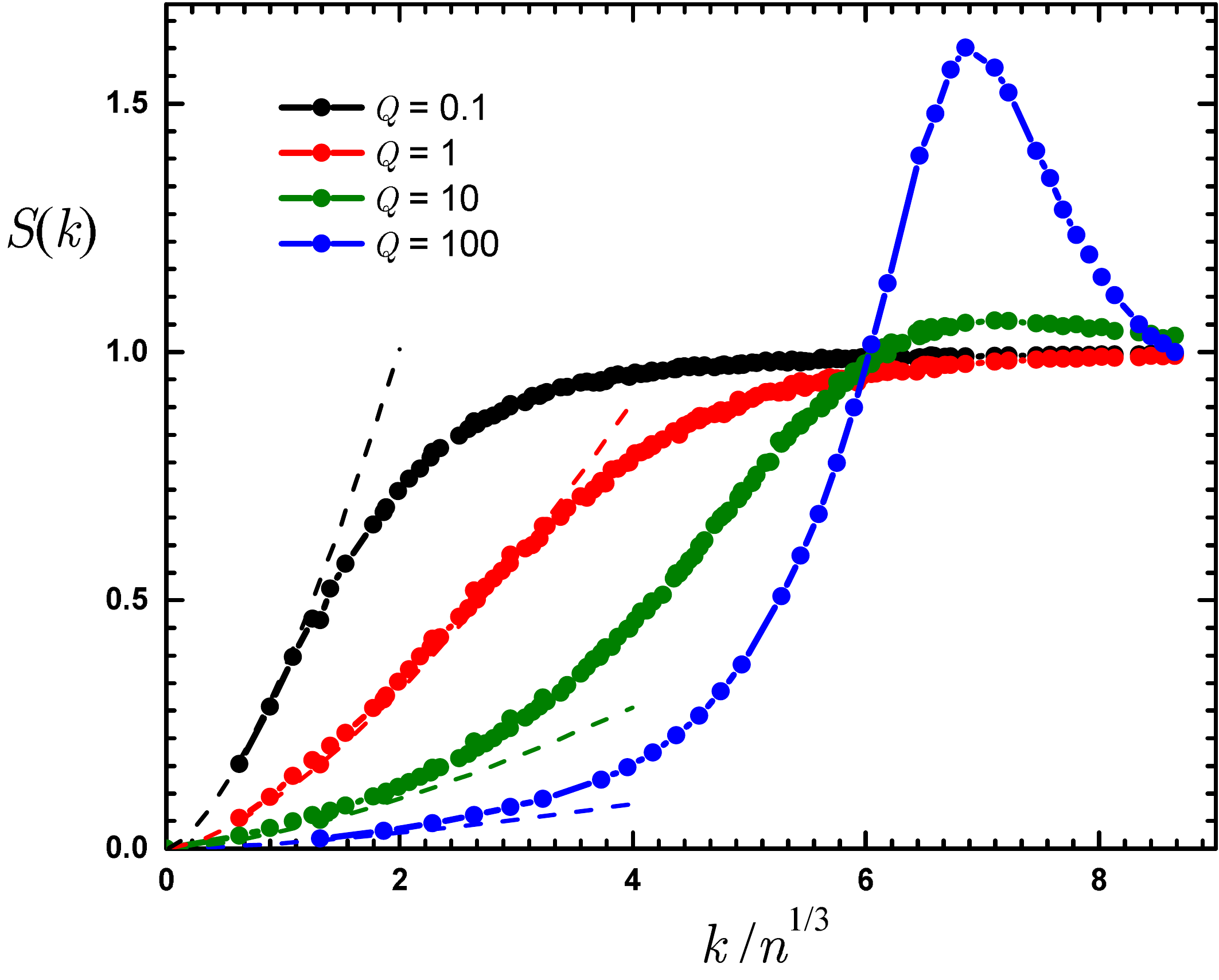}
\end{center}
\caption{
Static structure factor in the gas phase for different values of $Q=0.1; 1; 10; 100$ (from top to bottom).
Symbols, DMC data;
dashed lines, $k^{3/2}$ low momentum behavior with the coefficient of proportionality taken from Bogoliubov theory, Eq.~(\ref{Eq:Sk:Bog}).
}
\label{Fig:Sk}
\end{figure}

The pair correlations in momentum space can be quantified by the static structure factor $S(k)$ which is related to the Fourier transform of $g_2(r)$. The results for the gas phase are presented in Fig.~\ref{Fig:Sk}. In the weakly interacting regime, $Q \lesssim 1$, the static structure factor is a monotonous function which grows from zero at $k\to 0$ to its asymptotic large momentum value $S(k)=1$. Instead, in the regime of strong correlations, $Q\gg 1$, a peak forms. The height of the peak grows as $Q$ is increased which can be viewed as a precursor of the crystallization happening at the critical point $Q^*$.

Importantly, the small momentum part of the static structure factor is not linear, $S(k)\propto k$, as it happens in systems with a ``usual'' sound nor quadratic, $S(k)\propto k^2$, as in systems with a gap in the excitation spectrum\cite{Astrakharchik2016latt1D}. Instead, the plasmonic $k^{3/2}$ dependence is found. This plasmonic behavior is observed for different values of $Q$. In the Bogoliubov regime it is possible to calculate the corresponding prefactor,
\begin{eqnarray}
S(k) = \frac{|k|^{3/2}}{\sqrt{8\pi^2 Q n}}\;.
\label{Eq:Sk:Bog}
\end{eqnarray}

In the solid phase the static structure factor has a number of macroscopic peaks located at the corresponding momenta of the crystal lattice. The height of the peaks increases linearly with the number of particles signaling presence of the diagonal long-range order.

\subsection{Excitation spectrum and plasmons}

A peculiarity of the model described by Hamiltonian~(\ref{Eq:H}) is that the excitation spectrum is gapless but the low-lying excitations are not described by linear phonons but rather by plasmons with a square root dispersion relation. The Bogoliubov excitation spectrum~(\ref{Eq:Ebog}) provides an explicit expression in the limit of weak interactions.

In order to verify that the excitation spectrum follows the unusual plasmonic dispersion relation, we use the Feynman relation between the static structure factor $S(k)$ and the excitation spectrum $E(k)$,
\begin{eqnarray}
E(k) \leq \frac{\hbar^2k^2}{2mS(k)}\;.
\label{Eq:Feynman}
\end{eqnarray}
Estimation~(\ref{Eq:Feynman}) provides a rigorous upper bound for the lower boundary of the excitation spectrum. This expression becomes exact when the excitation spectrum is exhausted by a single type of excitation as happens in the limit of plasmons for $k\to 0$.

\begin{figure}[ht]
\begin{center}
\vskip 0 pt \includegraphics[clip,width=0.6\columnwidth]{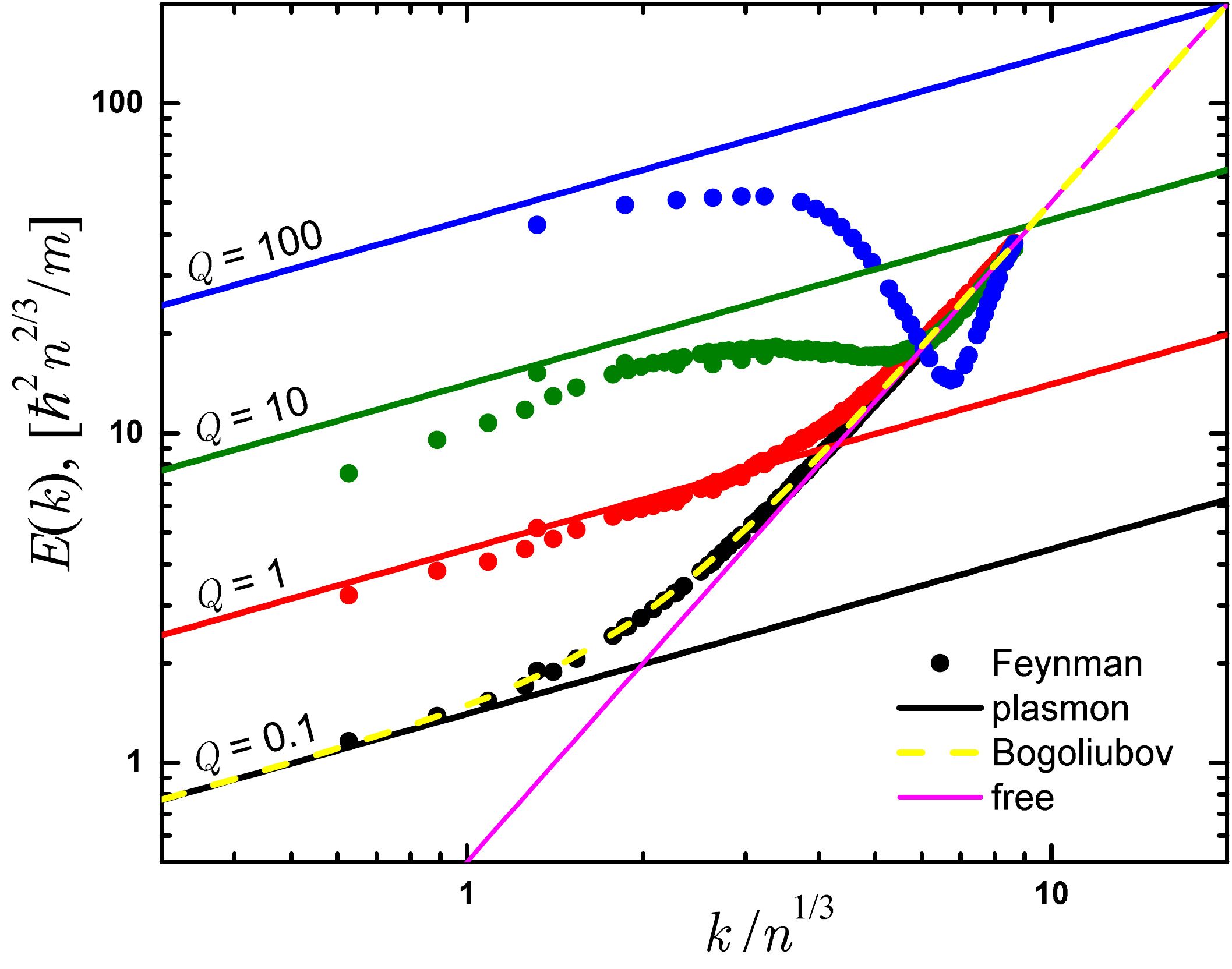}
\end{center}
\caption{
Upper bound for the excitation spectrum $E(k)$ on a double logarithmic scale as provided by Feynman relation~(\ref{Eq:Feynman}) by using the data presented in Fig.~\ref{Fig:Sk}.
Symbols,
Monte Carlo results;
solid thick lines, $k^{1/2}$ plasmon relation;
dashed line, Bogoliubov excitation spectrum~(\ref{Eq:Bogoliubov spectrum});
thin solid line, free particle $\hbar^2k^2/2m$ energy.
Units of $\hbar^2 n^{2/3}/m$ are used for the energy and $n^{1/3}$ for the momentum.
}
\label{Fig:Ek}
\end{figure}

We present the excitation spectrum as calculated from the Feynman relation in Fig.~\ref{Fig:Ek} for a number of characteristic values of the interaction parameter $Q$. The results are presented on a double logarithmic scale so that any power-law dependence appears as a straight line. We observe that for small momentum the excitation spectrum corresponds to plasmons with $\sqrt{k}$ dispersion relation. For the smallest considered interaction strength, $Q=0.1$, there is an excellent agreement with the Bogoliubov excitation spectrum given by Eq.~(\ref{Eq:Ebog}). For stronger interactions, the plasmonic behavior can be seen for small momenta, $E(k) \propto \sqrt{k}$. The coefficient of proportionality grows as $Q$ is increased reflecting stronger interactions in agreement both with the Bogoliubov theory and prediction Eq.~(\ref{Eq:E:HA:E(k)}) obtained within the harmonic crystal approximation. According to the Landau criterion for superfluidity, a condensate moving with a group velocity $V$ which is smaller than $\min_{k} E(k)/k$ (i.e. the speed of sound $c$ for linear spectrum with no roton) remains energetically stable. The ``speed of sound'' calculated as a first derivative of the dispersion relation diverges, $c\to\infty$, which means that according to the Landau criterion the homogeneous gas state is superfluid. For large momentum the excitation spectrum given by Eq.~(\ref{Eq:Feynman}) tends to the free particle spectrum, $E(k) = \hbar^2k^2/(2m)$, shown in Fig.~\ref{Fig:Ek} with a thin solid line. Close to the transition point, a roton minimum is formed at momenta similar to these where the Bragg crystal peak will be formed in the solid phase.

\subsection{Critical parameters}

It is illustrative to compare the properties of the system at the transition point to those of other quantum systems in order to check a possible universality. In particular the Lindemann parameter, the height of the peak in the static structure factor and the condensate fraction usually are of the same order as compared to other three-dimensional systems.

The Lindemann ratio, $\gamma = \sqrt{\langle r^2 \rangle} / d$, quantifies the strength of the mean square fluctuations of a particle close to the lattice site, $\langle r^2 \rangle$, compared to the nearest neighbor separation in the lattice ($d = a_0/\sqrt{2}$ for fcc lattice with elementary unit cell length $a_0$). The gas-solid phase transition is produced when the Lindemann ratio reaches a critical value which depends on the dimensionality and statistics of the system and to a lesser extent on the interaction potential.

\begin{table}[H]
\centering
\begin{tabular}{cccccccc}
\toprule
\textbf{Reference}& \textbf{Authors}& $\gamma$&$S_{peak}$(k)&$N_0/N$&\textbf{dim.}& \textbf{interaction} & \textbf{method}\\
\midrule
---&
{\bf present work}&
{\bf 0.24(1)}&
{\bf1.63(5)}&
{\bf0.008(2)}&
{\bf 3D}&
$1/r^2$&
{\bf DMC}\\
\cite{Cazorla09} & C. Cazorla et al.
&0.26(1)		&---		&---	&3D& $^4$He&	DMC\\
\cite{PhysRevB.42.228}& S. A. Vitiello et al.
&0.23(1)		&1.55		&---	&3D& $^4$He&	GFMC\\
\cite{PhysRevLett.60.1970}	& S. A. Vitiello et al.&
0.23(1)		&---		&---	&3D& $^4$He&	VMC\\
\cite{CeperleyChesterKalos78}& D. Ceperley et al.
&0.28(2)&---		&---	&3D&Yukawa&DMC\\
\cite{PhysRevLett.64.1529}&A. R. Denton et al.
&0.27(1)		&---		&---	&3D& hard core&	DFT\\
\cite{PhysRevLett.76.4572} & M. D. Jones and D. M. Ceperley
&0.27(1)		&---		&---	&3D& electrons&	PIMC\\
\cite{PhysRevB.18.3126}& D. Ceperley
&0.29(1)		&---		&---	&2D& electrons&	VMC\\
\cite{PhysRevB.38.2418}& P. A. Whitlock et al.
&0.254(2)	&1.70(2)		&---	&2D& $^4$He&	GFMC\\
\cite{PhysRevB.42.8426}& L. Xing
&0.279(1)	&1.54(2)		&---	&2D& hard core&	DMC\\
\cite{PhysRevB.48.411} & W. R. Magro and D. M. Ceperley
&0.245(15)       &---		&---	&2D& Yukawa&	DMC\\
\cite{PhysRevLett.73.826}& W. R. Magro and D. M. Ceperley
&0.24(1)		&---		&---	&2D& 1/$r$ charges&	DMC\\
\cite{Astrakharchik07a}& G. E. Astrakharchik et al.
	&0.230(6)	&1.70(3)		&0.014(2)&2D& dipoles&	DMC\\
\bottomrule
\end{tabular}
\caption{Literature overview for the typical values of the Lindemann parameter $\gamma$ (solid phase),
height of the peak of the static structure factor $S_{peak}(k)$ and
condensate fraction $N_0/N$ (gas phase) at the critical point of zero-temperature gas(liquid)-solid phase transition.
}
\end{table}

The Lindemann ratio is approximately constant along the transition line  and is relatively independent of the types of interaction potential and the crystal packing. For example, the phase diagram for the Yukawa potential is governed by two parameters and the explicit calculation of the transition line\cite{Osychenko2012yukawa} is very close to the prediction based on a constant value of $\gamma = 0.28(2)$\cite{CeperleyChesterKalos78}. Thus, the value of the Lindemann ratio provides important information on the phase transition location and its value can be compared with what is observed in other systems. The Lindemann ratio is also approximately constant in classical systems at the transition, although its typical values  $0.1 \lesssim \gamma \lesssim 0.15$ are smaller compared to the quantum systems\cite{Yaoqi2002}

Table~1 summarizes the value of the Lindemann parameter at the zero-temperature phase transition point in a number of different systems in three (3D) and two (2D) dimensions. It can be noted that even if the interaction potentials might be very different, including long-range ones, the actual value of the Lindemann ratio is limited to a rather narrow range, $0.23<\gamma<0.29$. We find that at the transition the Lindemann parameter is equal to $\gamma = 0.24(1)$. In other words the present system falls into the same class of quantum phase transitions. Instead, classical systems have typically a smaller value of the Lindemann ratio at gas-solid transition\cite{CazorlaBoronat2017RMP}.

It might be noted that also the gas phase has same parameters which are approximately constant at the critical point, which are the height of the peak in the static structure factor and the condensate fraction. We find that the condensate fraction is $0.8(2)\%$ at the transition point and the height of the peak of the static structure factor is $1.63(5)$.

\subsection{Universal scaling properties}

A number of system properties are universal in that they can be mapped to the properties of an ideal Fermi gas and are defined by a single function, which can be formally introduced as the dependence of the Bertsch parameter $\xi$ on the interaction parameter $Q$.

The total energy per particle can be generally written as
\begin{eqnarray}
\frac{E}{N} = \xi(Q) \frac{\hbar^2N^{2/3}}{mV^{2/3}} + \frac{1}{2}C NQ \frac{\hbar^2}{mV^{2/3}}\;,
\label{Eq:scaling:E}
\end{eqnarray}
where the energy unit, apart from a numerical factor, coincides with the Fermi energy and we explicitly add the energy of the jellium background. The chemical potential can be calculated from the total energy~(\ref{Eq:scaling:E}) according to $\mu = \partial E / \partial N$ resulting in
\begin{eqnarray}
\mu = \frac{5}{3}\xi(Q) \frac{\hbar^2N^{2/3}}{mV^{2/3}}+ C NQ \frac{\hbar^2}{mV^{2/3}}\;.
\label{Eq:scaling:mu}
\end{eqnarray}

The Bertsch parameter reported in Fig.~\ref{Fig:E} is obtained by subtracting the diverging jellium contribution from the energy per particle. The same is true for the chemical potential, once the jellium contribution is subtracted, the remaining term follows the ideal Fermi gas scaling. While the mean-field (jellium) contribution can be conveniently subtracted from the energy and the chemical potential as the potential energy can be calculated with a certain offset, the jellium contribution to the compressibility becomes crucial to the small-momentum properties.
The pressure
$P = - \left.\partial E/\partial V\right|_{N}$
is the sum of the ``Fermi gas'' contribution
$\frac{2}{3}\xi(Q)\hbar^2n^{5/3}/m$
and the diverging ``jellium'' contribution $\propto N^{1/3}Q \hbar^2n^{5/3}/m$.
The latter term has a dramatic effect on the compressibility $\kappa$ which at zero temperature can be calculated as $\kappa^{-1}= (V/m^2)\left.\partial^2E / \partial N^2\right|_V$. Due to the jellium contribution its value vanishes in the thermodynamic limit, $\kappa\to 0$. The relation between compressibility and the speed of sound, $c^2 = mn/\kappa$, results in a diverging value of the speed of sound $c$, seen in Fig.~\ref{Fig:Ek} as an infinite slope of the plasmonic excitation spectrum for small momentum. A related effect was observed in classical simulations of two-dimensional charges in a large trap\cite{PhysRevB.49.2667} where long-range interactions resulted in vanishing compressibility leading to a larger concentration of charges at the border of the trap.

The potential energy $E_{pot}$ can be obtained by using the Hellmann - Feynman theorem by noting that $Q$ enters in the Hamiltonian~(\ref{Eq:H}) only in the interaction energy, so that the potential energy per particle is obtained by differentiating the Hamiltonian with respect to $Q$ and exchanging the order of the derivative and averaging
\begin{eqnarray}
\frac{E_{pot}}{N}
= \frac{Q}{N}\left\langle \frac{dH}{dQ} \right\rangle
= \frac{Q}{N} \frac{d\left\langle H\right\rangle}{dQ}
= \frac{Q}{N} \frac{dE(Q)}{dQ}
=
Q\xi'(Q) \frac{\hbar^2N^{2/3}}{mV^{2/3}} + \frac{1}{2}C NQ \frac{\hbar^2}{mV^{2/3}}\;.
\label{Eq:scaling:Epot}
\end{eqnarray}
The potential energy per particle~(\ref{Eq:scaling:Epot}) diverges in the thermodynamic limit ($N\to\infty$ taken with $N/V-const$) due to the long-range nature of interactions. At the same time, the kinetic energy per particle remains finite
\begin{eqnarray}
\frac{E_{kin}}{N}
=
\frac{E}{N}
-
\frac{E_{pot}}{N}
=
(\xi(Q)-Q\xi'(Q)) \frac{\hbar^2N^{2/3}}{mV^{2/3}},
\label{Eq:scaling:Ekin}
\end{eqnarray}
and scales as $\hbar^{2/3}n^{2/3}/m$, similar to the kinetic energy of an ideal Fermi gas. The kinetic energy appears due to a non-linear dependence of the Bertsch parameter on $Q$. In the limiting case of a classical crystal, $Q\to\infty$, the Bertsch parameter becomes linear according to Eq.~(\ref{Eq:E:fcc}) so that the leading contribution to the energy of a classical crystal comes from the potential energy. The beyond-mean field energy per particle is given by Eq.~(\ref{Eq:E:BMF:Q}), consequently
\begin{eqnarray}
\label{Eq:E:BMF:Epot}
E^{BMF}_{pot}& =& -\frac{5}{3}C_{BMF} N Q^{5/3} \frac{\hbar^2n^{2/3}}{m}\;,\\
E_{kin}& =& \frac{2}{3}C_{BMF} N Q^{5/3} \frac{\hbar^2n^{2/3}}{m}\;,
\label{Eq:E:BMF:Ekin}
\end{eqnarray}
for small $Q$, $E_{kin}/|E_{pot}^{BMF}| = 2/5$.

\section{Considerations for experimental realization}

Although no explicit realization of the system under study is known to us, a number of closely related systems already exist or can be experimentally realized in the near future.

There is a close analogy between the statistical properties of zero-temperature one-dimensional quantum gases and terraces on crystal surfaces\cite{Einstein91,Gebremariam04,Einstein2007,Jaramillo13}.
Typically, the crystal surface is covered by molecular layers of the same height (terraces).
Different terraces are separated by steps at which the elevation is changed by the height of a single elementary cell.
The border of a step seen from above draws a trajectory on a two-dimensional $(x,y)$ surface and can be interpeted as a word line of a quantum 1D line in the $(x,\tau)$ plane where imaginary time $\tau$ plays the role of another dimension.
Energetically it is not favorable to have a step of double height so that the steps (and equivalently the world lines) do not cross each other, creating an analogy with the Pauli exclusion principle.
The mass for quantum particles is than mapped to the step stiffness $\beta$ while thermal energy $k_BT$ replaces the Planck's constant $\hbar$\cite{Einstein2007}.
The elastic repulsion when meandering is modest (variation in $x$ is small compared to the average distance between steps $\ell$) leads to energy $A T^2/ (\beta \ell^2)$ where $A$ is a constant\cite{Gebremariam04,Jaramillo13} fixed by the material.
The mapping to the quantum system results in particles interacting via an inverse square potential.
The dimensionless parameter of the problem, $\tilde A = A \beta / (k_BT)^2$  is equivalent to $Q$ in our terminology and it changes in a wide range, $0.1 \lesssim A \lesssim 100$\cite{Einstein2007} in real materials.
The mapping with the Calogero-Sutherland model turns out to be useful for comparison of the energy and density-density (step-step) correlation function\cite{Einstein91,Gebremariam04,Einstein2007,Jaramillo13}.

Another class of systems where the inverse square interaction potential can be experimentally realized are ions with controllable spin-spin interactions\cite{Grass2014} relevant for the creation of gates needed in a quantum computer.
It was experimentally demonstrated that power-law interactions, $V(r) \propto 1/r^\alpha$, with $0\leq \alpha \leq 3$ can be  induced in a two-dimensional triangular crystal lattice of hundreds of particles\cite{Britton2012}, including ``monopole–dipole'' $\alpha=2$ case corresponding exactly to the inverse square interactions.
One-dimensional ion chains with power-law interactions with $0.75 \leq \alpha\leq 1.75$ were experimentally realized in Ref.~\cite{Jurcevic2014} and observed by different behavior of the light-cone as a function of $\alpha$\cite{Hauke2013}.
It is rather probable that the required interaction potential will be also created in three dimensional ion lattices.

The inverse-square interaction is relevant for Rydberg atoms\cite{Bendkowsky2009} and polar molecules\cite{PhysRevLett.73.2436} as well as polymer physics\cite{MarinariParisi1991}.
The renormalization-group theory was used to predict properties in arbitrary number of dimensions\cite{KolomeiskyStraley1992,KolomeiskyStraley1992b,Kolomeisky1994}.
Also the relation between the Laughlin state and the wave function of Calogero-Sutherland model was shown in Refs.~\cite{Azuma1994, PhysRevB.52.13742, Feinberg2005}.

\section{Conclusions}

We have studied the ground state properties of a three-dimensional quantum system with particles interacting via an inverse squared pair potential of strength $Q$. Its intrinsic property is that it is scale-free with the density being the only parameter providing a length scale. The system properties can be divided into two categories. The first one corresponds to mean-field quantities which are sensitive to the long-range nature of the interaction potential and which diverge in the thermodynamic limit, including total energy, potential energy, speed of sound, etc. The second category describes beyond-mean field properties which remain finite in the thermodynamic limit, including the kinetic energy, excitation spectrum, condensate fraction, Lindemann ratio, etc. The properties belonging to the second category can be naturally expressed in terms of the Fermi momentum and the Fermi energy. Diffusion Monte Carlo method is used to calculate numerically the ground-state properties for a wide range of parameters, $0.1<Q<1000$. The guiding wave function is constructed from a power-law two body solution at short distances and a ``plasmonic'' long range Jastrow tail. We demonstrate that this system possesses a gas-solid phase transition. The energies of the gas and solid phases are separately calculated. We estimate the critical value of the interaction parameter as $Q^* = 94(5)$. Notably, the density dependence of the solid phase is still that of an ideal Fermi gas.

For weak interactions, $Q\to 0$, we develop a perturbative approach based on Bogoliubov theory, complemented with a jellium model which is used to remove the mean-field divergence of the energy in the thermodynamic limit. For strong interactions, $Q\to\infty$, we use the harmonic crystal theory. A peculiarity emerging from the long-range nature of the interactions is that the low-lying excitations are gapless plasmons with a square root dispersion relation which is demonstrated within Bogoliubov theory and the harmonic crystal approach. According to the Landau criterion the homogeneous gas is superfluid at zero temperature.

We validate the results derived within the Bogoliubov theory for small $Q$ and the harmonic approximation for large $Q$. In particular, we show that the energy correction is well reproduced by beyond-mean-field terms for small values of $Q$ and approaches the energy of a classical crystal when $Q\to\infty$. We verify that condensate depletion is caused by quantum fluctuations in the limit of weak interactions and becomes very small in the gas phase close to the transition point. A number of characteristic examples of the pair-distribution function $g_2(r) $are reported across the gas phase. We show that for small distances it follows a non-analytic law, $g_2(r)\propto |r|^{2\lambda}$. The correlations in the system are quantified by the static structure factor. The excitation spectrum is approximated by using the Feynman relation and shows the plasmonic dispersion for small momenta. In the limit of weak interactions it coincides with the Bogoliubov excitation spectrum.

The unitary scaling with the density permits us to find intrinsic relations between different thermodynamic quantities. In particular, by using the Bertsch parameter $\xi(Q)$ and its derivatives, it is possible to predict the energy, chemical potential and pressure. One of the main results is the prediction for the Bertsch parameter $\xi(Q)$ as calculated from the ground state energy, $E/N = \xi(Q)\hbar^2n^2/(2m)$, for which we provide a Pad\'e approximant. By using Helmann-Feynman theorem we also provide explicit expressions for the potential and kinetic energy. We verify such predictions by using Monte Carlo data which demonstrates the internal consistency of the obtained results.

As a consequence of the scaling properties, the dynamics is equivalent to that of an ideal Fermi gas and this model can be interpreted as a realization of a unitary regime in a Bose system. Importantly, in our case the particles stay in a genuine ground state and not in a metastable state, as instead happens in experiments with short-ranged Bose gases.

\vspace{6pt}

\acknowledgments{
We thank Russell Bisset for reading the Manuscript.
The research leading to these results received funding from the MICINN (Spain) Grant No. FIS2014-56257-C2-1-P. The Barcelona Supercomputing Center (The Spanish National Supercomputing Center - Centro Nacional de Supercomputaci\'on) is acknowledged for the provided computational facilities (RES-FI-2018-1-0005).
The work of Y.E.L. was supported by the Program for Basic Research of the National Research University Higher School of Economics.
}



\abbreviations{The following abbreviations are used in this manuscript:\\

\noindent
\begin{tabular}{@{}ll}
DMC & diffusion Monte Carlo\\
HA & Harmonic approximation\\
\end{tabular}}

\appendixtitles{no} 
\appendixsections{multiple} 

\reftitle{References}
\externalbibliography{yes}


\end{document}